# A tidal disruption event in the nearby ultra-luminous infrared galaxy F01004-2237

C. Tadhunter[1], R. Spence[1], M. Rose[1], J. Mullaney[1], P. Crowther[1]

[1]Department of Physics & Astronomy, University of Sheffield, Hounsfield Road, Sheffield S3 7RH

**Tidal disruption events (TDEs), in which stars are gravitationally disrupted as they pass close to the supermassive black holes in the centres of galaxies[1], are potentially important probes of strong gravity and accretion physics. Most TDEs have been discovered in large-area monitoring surveys of many 1000s of galaxies, and the rate deduced for such events is relatively low: one event every $10^4 - 10^5$ years per galaxy[2-4]. However, given the selection effects inherent in such surveys, considerable uncertainties remain about the conditions that favour TDEs. Here we report the detection of unusually strong and broad helium emission lines following a luminous optical flare ($M_V$<-20.1 mag) in the nucleus of the nearby ultra-luminous infrared galaxy F01004-2237. The particular combination of variability and post-flare emission line spectrum observed in F01004-2237 is unlike any known supernova or active galactic nucleus. Therefore, the most plausible explanation for this phenomenon is a TDE– the first detected in a galaxy with an ongoing massive starburst. The fact that this event has been detected in repeat spectroscopic observations of a sample of 15 ultra-luminous infrared galaxies over a period of just 10 years suggests that the rate of TDEs is much higher in such objects than in the general galaxy population.**

Ultra-luminous infrared galaxies (ULIRGs: $L_{IR} > 10^{12}$ $L_\odot$)[5] represent the peaks of major, gas-rich galaxy mergers in which the merger-induced gas flows concentrate the gas into the nuclear regions, leading to high rates of star formation and accretion onto the central supermassive black holes. The nearby ULIRG F01004-2237 (RA: 01h 02m 50.007s and Dec: -22d 21m 57.22s (J2000); z=0.117835) was observed using deep spectroscopic observations in September 2015 as part of a study of 15 ULIRGs to examine the importance of the warm gas outflows driven by active galactic nuclei (AGNs) in such objects[6]. Many of its properties are typical of local ULIRGs, including relatively modest total stellar and supermassive black hole masses ($M_* = 1.9 \times 10^{10}$ $M_\odot$; $M_{bh} = 2.5 \times 10^7$ $M_\odot$)[7,8], and evidence for AGN activity in the form of blue-shifted high ionization emission lines[6]. However, it is unusual in the sense that it is one of the few ULIRGs in which Wolf-Rayet features have been detected at optical wavelengths[9], indicating the presence of a population of $\sim 3 \times 10^4$ Wolf-Rayet stars with ages 3-6 Myr (supplementary information). Also, unlike most ULIRGs for which the central starburst regions are heavily enshrouded in dust, this source has a compact nucleus that is barely resolved in optical and UV observations with the Hubble Space Telescope (HST) and has been attributed to a population of stars that is both young ($< 10$ Myr) and massive ($\sim 3 \times 10^8$ $M_\odot$)[10]. Together, these features suggest that we have an unusually clear view of the nuclear star forming regions in F01004-2237.

Figure 1 compares the optical spectra of F01004-2237 taken in 2015 with earlier spectra taken in 2000. In common with all the other published optical spectra of the source taken between 1995 and 2005 (Supplementary Information), those obtained in 2000 show emission lines characteristic of a composite of regions photoionized by hot stars and AGN, as well as a blend of NIII and HeII emission lines at ~4660Å that is characteristic of Wolf-Rayet stars of type WN[9]. However, the 2015 spectra are markedly different: the ~4660Å feature is a factor of 5.6+/-1.1 stronger in flux compared with 2000, and the blend has developed broad wings that extend up to ~5,000 km/s to the red of the centroid of the narrow HeIIλ4686 component; the HeIλ5876 line is also notably stronger (by a factor 3.7+/-0.2); and new HeI emission features have appeared at 3889, 4471, 6678 and 7065Å. In contrast, the broad, blue-shifted forbidden lines associated with the AGN-induced outflow have not varied significantly between the two epochs (Supplementary Information), whereas the Hβ line has increased by a factor of 1.52+/-0.12. If the broader component of the blend at 4660Å detected in the 2015 spectrum is attributed to HeIIλ4686, the ratio of the flux of this broad line to that of the broader Hβ component is HeIIλ4686/Hβ=1.82+/-0.09; however, the ratio is much larger if we only consider the component of Hβ that has changed between the two epochs: HeIIλ4686/Hβ=5.3+/-0.85. Similarly, we derive HeIλ5876/Hβ=0.46+/-0.04 and HeIλ5876/Hβ=1.34+/-0.23 when comparing the broad HeIλ5876 flux to the total and variable broad Hβ component fluxes respectively. In comparison, typical quasars have HeIIλ4686/Hβ~0.02 and HeIλ5876/Hβ~0.09 for their wavelength-integrated broad emission line fluxes[11].

Alerted by our spectra to the possibility of an unusual transient event in the nucleus of F01004-2237, we examined the Catalina Sky Survey (CSS)[12] database for evidence of variability in its optical continuum over the period 2003 – 2015. The resulting V-band light curve for F01004-2237 is presented in Figure 2, where it is compared with those of the other 14 ULIRGs in our spectroscopic sample. Due to their low spatial resolution, the Catalina measurements include a substantial fraction of the total light of each galaxy, not just that of the nucleus. Whereas the light curves of all the other ULIRGs are flat within +/-0.1 magnitudes, that of F01004-2237 showed a significant spike in 2010, when it was 0.45+/-0.02 magnitudes (a factor of 1.51+/-0.03) brighter than the average of the four earliest epochs.

A supernova origin for the phenomenon observed in F01004-2237 is ruled out by the fact that the light curve and post-flare spectrum are unlike any known supernova. Although high rates of supernovae are expected in ULIRGs because of their high star formation rates (>100 $M_\odot$ yr$^{-1}$), and a rate of 4+/-2 yr$^{-1}$ has been measured for the closest ULIRG – Arp220 – using radio observations[13], most core-collapse supernovae would not be sufficiently bright to detect in the integrated-light CSS observations of F01004-2237. The peak luminosity of the flare in F01004-2237 ($M_v$<-20.1 mag) approaches that of super-luminous supernovae, which are orders of magnitude less common than typical core-collapse supernovae.

Given the prior evidence for an AGN in F01004-2237, it is also important to consider whether the optical flare and spectral changes observed in F01004-2237 fall within the range of observed AGN activity. High amplitude (more than a factor of a ten) flares are not unprecedented in AGN. However, they are rare in the type of radio-quiet AGN represented by F01004-2237 that lack powerful, synchrotron-emitting jets.

Considering the class of "changing-look" AGN in which strong broad emission lines and non-stellar continuum have appeared in optical spectra after a period of apparent quiescence, it is notable that the broad HeI and HeII lines are not unusually strong in the high-state spectra of such objects[14,15] (see Supplementary Information for details). As far as we are aware, the variability observed in F01004-2237, in which the broad helium emission lines dominate the high-state spectrum, is without precedent for an AGN. Moreover, the HeIIλ4686/Hβ and HeIλ5876/Hβ ratios measured for the broad, variable emission lines in F01004-2237 are significantly higher than those measured for even the innermost, highest ionization zones of typical AGN broad line regions (HeIIλ4686/Hβ~1; HeIλ5876/Hβ~0.5 – 0.6)[16,17], as represented by extreme red and blue wings of the emission line profiles.

Although not typical of AGN, unusually strong and variable broad HeI and HeII lines have been observed in some tidal disruption events (TDEs)[18-20]. Therefore, the most plausible explanation for the unusual properties of F01004-2237 is a TDE that took place ~5 years before the 2015 spectroscopic observations. The absolute V-band luminosity of the peak of the transient event ($M_v$ <-20.1 mag) is characteristic of TDEs[19]. However, the flare in F01004-2237 is unusually prolonged compared with typical TDEs[18], with the light curve appearing to flatten at late times rather than follow the $(t/t_{peak})^{-5/3}$ decline predicted by theory ($t_{peak}$ is the time taken for the flare to reach peak brightness following the disruption of the star). In order to explain the slow decline in the light curve of F01004-2237 for the first three years following the peak brightness, a relatively long $t_{peak}$ would be required ($t_{peak}$~1 yr). Assuming that the star was disrupted by a single super-massive black hole (SMBH) of mass $M_{bh} = 2.5 \times 10^7$ $M_\odot$,[8] this in turn would favour a relatively high polytropic index ($\gamma$~5/3),[21] corresponding to a low mass star ($M_* \leq 0.3$ $M_\odot$) with a fully convective envelope. The prolonged nature of the continuum flare might also help to explain why we observe emission lines from the debris 5 years after the event, whereas in some other well-observed TDEs with shorter $t_{peak}$ the emission lines had become undetectable on such timescales. Explaining the apparent flattening of the light curve ~5 yr after the peak – albeit based on only one photometric point with a relatively large error bar – is more challenging, but it is notable that flattening or re-brightening has recently been detected in the light curves of some other TDEs[22,23].

The cause of the strong helium lines detected in the optical spectra of TDEs is the subject of debate. Recently it has been proposed that, even in the case of solar abundances, the ratio of the helium to hydrogen lines might be substantially enhanced if the emitting gas is both matter-bounded and has a sufficiently high density that the Balmer lines of hydrogen are optically thick[16,24]. In the case of F01004-2237, this explanation is favoured over the alternative that the He/H abundance ratio is substantially super-solar in the tidal debris[25]. This is because such an enhanced helium abundance would require the disrupted star to be sufficiently massive (>10 $M_\odot$) that it had converted much of its hydrogen to helium over the <10 Myr lifetime of the nuclear star cluster; the tidal disruption of such a massive star is expected to be extremely rare for a typical initial mass function.

The detection of one event in a sample of just 15 ULIRGs over a period of ~10 years suggests that the rate of TDEs in such objects is orders of magnitude higher than the $10^{-5} – 10^{-4}$ TDE yr$^{-1}$ galaxy$^{-1}$ deduced for the field galaxy population,[2-4,20] and perhaps as high as $10^{-2}$ TDE yr$^{-1}$ galaxy$^{-1}$. This is also higher than the TDE rate recently deduced for the population of post-starburst galaxies (~$10^{-3}$ TDE yr$^{-1}$ galaxy$^{-1}$)[26]. However, considering that we have an unusually clear view of the nucleus in F01004-2237, whereas in the remainder of our sample the nuclear regions are likely to be heavily obscured by dust at optical wavelengths, the true rate of TDEs in the ULIRG population could be higher still (~$10^{-1}$ TDE yr$^{-1}$ galaxy$^{-1}$).

Several mechanisms could enhance the rate of TDEs in ULIRGs, including concentrated nuclear star formation leading to high densities of stars close to the central SMBHs[27], the formation of close black hole binaries comprising the SMBHs of the progenitor galaxies[28], and black hole recoils that might follow the coalescence of such binaries[29]. We note that the disruptive effect of a SMBH binary on the debris stream of the TDE[30] might also help to explain the fact that the flare observed in F01004-2237 is unusually prolonged.

The simultaneous detection of a TDE, a massive young stellar population, and an AGN in the nucleus of a ULIRG provides dramatic evidence of the close proximity of star formation and growing SMBHs at the centres of starburst galaxies. The TDE flare in F01004-2237 has required the consumption of a total mass of $0.02 - 0.61$ M$_\odot$ by its SMBH (see Methods). This corresponds to an average mass accretion rate of $2\times10^{-4} < \dot{M} < 6.1\times10^{-2}$ M$_\odot$ yr$^{-1}$, assuming a TDE rate in the range $10^{-2} < R_{TDE} < 10^{-1}$ yr$^{-1}$. If this rate were maintained for the ~100 Myr timescale typical of starbursts in ULIRGs, the SMBH in F01004-2237 would grow in mass by $2\times10^4 - 6.1\times10^6$ M$_\odot$ (0.1 – 25%) due to TDEs alone. Although not sufficient to trigger a luminous, quasar-like episode of AGN activity, the integrated photoionizing effect of the frequent TDE flares would be capable of sustaining lower-level LINER- or Seyfert-like narrow emission line activity in ULIRGs in periods of relative quiescence, when the rates of direct gas accretion onto the SMBH are low.

Correspondence and requests for materials should be addressed to Clive Tadhunter (c.tadhunter@sheffield.ac.uk).

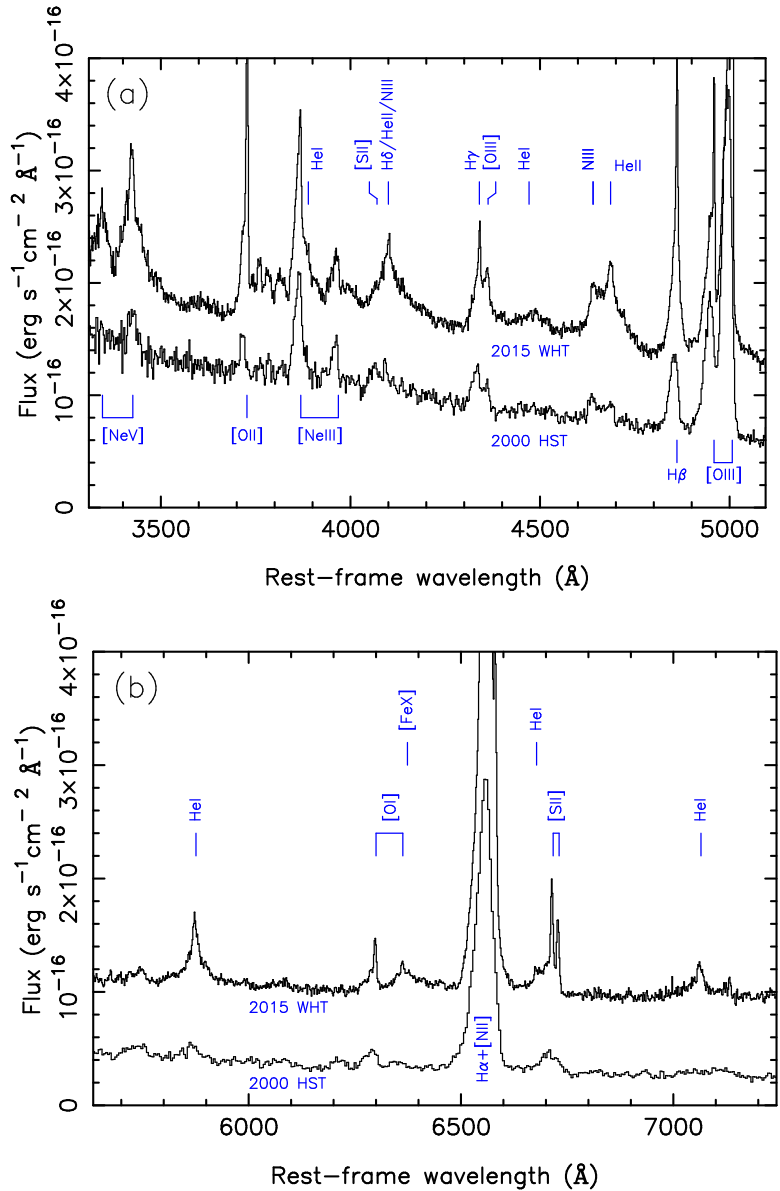

**Figure 1.** Comparison of optical spectra of F01004-2237 taken in September 2015 using the ISIS spectrograph on the William Herschel Telescope (WHT) with those taken in September 2000 using the STIS spectrograph on the Hubble Space Telescope (HST): (a) tcomparison of the blue spectra; (b) comparison of the red spectra. Details of all the spectroscopic observations taken at this and other epochs are given in the Supplementary Information. The most probable line identifications are indicated. The narrowest emission line components visible in the spectra are likely to be emitted by the extended regions around the nucleus and are therefore more prominent in the WHT spectra that were taken with a wide spectroscopic slit (1.5 arcsec) but are weak in the HST spectra taken with a narrower slit (0.2 arcsec). Note the detection of broad HeII and HeI emission line components in 2015 that are not visible in the Hβ and Hγ Balmer lines.

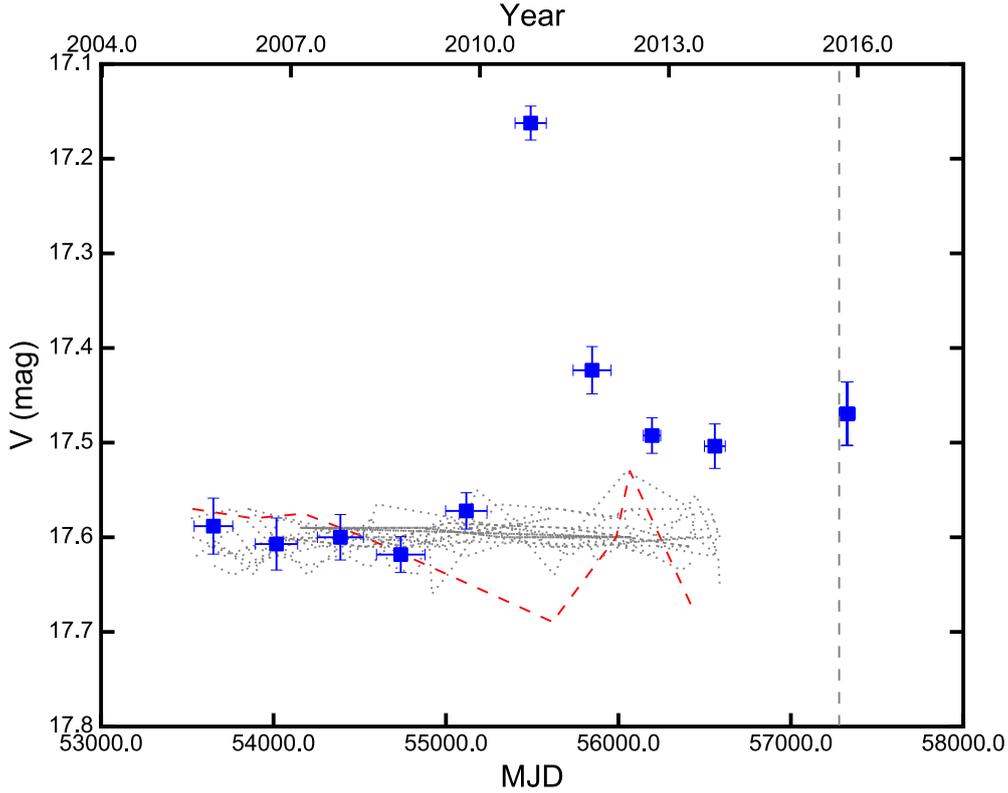

**Figure 2.** Catalina Sky Survey (CSS)[12] light curves for F01004-2237 (solid blue points) and the other 14 sources in our spectroscopic sample (black dotted lines and red dashed line). All the light curves have been shifted to have the same mean V-band magnitude as F01004-2237 in the first three epochs. Note that, whereas F01004-2237 has shown a substantial flare in its V-band brightness ($\Delta m_v$=0.45+/-0.02 mag) over the ~10 years of the survey, none of the other sources have shown similar flares. In the majority of the ULIRGs with type 2 AGN spectra, the level of variability is $|\Delta m_v|$ < 0.07 magnitudes (black dotted lines). Although the one type 1 AGN in our sample (red dashed line) shows evidence for a higher level of variability – as expected for an AGN in which the nucleus is directly visible – it is notably less variable than F01004-2237. The vertical dashed line indicates the date of the September 2015 WHT observations. The blue points represent the mean magnitudes obtained by averaging all the measurements (between 12 and 37 measurements per epoch) in the CSS database available for each observation epoch in which F01004-2237 was observed. The horizontal bars indicate the range of dates covered by each observation period, whereas the vertical bars indicate the standard errors in the means.

**Acknowledgements**. The William Herschel Telescope is operated on the island of La Palma by the Isaac Newton Group in the Spanish Observatorio del Roque de los Muchachos of the Instituto de Astrofisica de Canaria. Based on observations made with the NASA/ESA Hubble Space Telescope, obtained from the Data Archive at the Space Telescope Science Institute, which is operated by the Association of Universities for Research in Astronomy, Inc., under NASA contract NAS 5-26555, these observations are associated with program #8190. This project made use of data obtained by the Catalina Sky Survey[12]. CT, RS, MR and PC acknowledge financial support from the UK Science & Technology Facilities Council. We thank Justyn Maund for useful discussions about the possibility of a supernova origin for the flare.

**Author contributions.** CT and RS led the project and the scientific interpretation of the data, and CT wrote the text of the paper. MR extracted the Catalina Sky Survey light curves and contributed to the general interpretation of the emission line spectra. JM and PC contributed equally to the analysis and interpretation of the results.

## Methods

We have estimated the absolute magnitude of the peak of the flare in F01004-2237 by assuming a cosmology with $H_0$=73.0 km s$^{-1}$ Mpc$^{-1}$, $\Omega_m$=0.27, $\Omega_\lambda$=0.73, which results in a luminosity distance of $D_L$=523 Mpc for the redshift of F01004-2237 (z=0.117835). The SED of the flare is unknown, but if we assume that it follows the Rayleigh-Jeans tail of a hot black body (T>20,000 K), the K-correction is 0.24 magnitudes. Applying this K-correction and a Galactic extinction correction of $A_V$=0.05 mag, we derive an absolute magnitude for the peak of the flare of $M_V$=-20.1 mag. However, this is likely to represent a lower limit on the luminosity, since we have not corrected for intrinsic dust extinction.

To calculate the bolometric luminosity associated with the flare it is necessary to assume a bolometric correction factor (BCF) to convert between the V-band monochromatic luminosity and the bolometric luminosity. This BCF depends on the (unknown) SED of the flare. Assuming that the SED of the flare in F01004-2237 is similar to that of other TDEs with multi-wavelength photometry and follows a black body with temperature in the range 10 – 50 kK, the BCF will be in the range $1.75 < L_{bol}/\nu L_V < 60$. For comparison, typical AGN have $L_{bol}/\nu L_V \sim 8$. Considering the full range of likely black body temperatures, the peak bolometric luminosity of the TDE flare in F01004-2237 falls in the range $4\times10^{43} < L_{bol}(peak) < 1.4\times10^{45}$ erg s$^{-1}$, and performing a simple trapezium rule integration of the light curve, the total energy associated with the flare up to the end of 2015 was in the range $3\times10^{51} < E_{flare} < 1.1\times10^{53}$ erg.

The mass consumed by the black hole in order to produce the flare is $M_{flare}=E_{flare}/(c^2\eta)$, where $\eta$ is the efficiency. Therefore, for a typical SMBH accretion disk efficiency of $\eta$=0.1, the total mass consumed so far to produce the TDE flare observed in F01004-2237 is in the range $0.02 < M_{flare} < 0.61\ M_\odot$.

**Data availability:** the data used to make the photometric light curve presented in Figure 2 are available from the CSS data release 2 website (http://nunuku.caltech.edu/cgi-bin/getcssconedb_release_img.cgi); the data that support Figure 1 within this paper and other findings of this study are available from the corresponding author upon reasonable request.

# Supplementary information

## 1. Parent sample

F01004-2237 is one of 15 local ULIRGs displaying signs of optical AGN activity for which we have taken optical spectra at two epochs: 2005-2006 and 2013-2015. This sample is itself a subset of a larger sample that comprises all 22 ULIRGs in the 1Jy sample of Kim et al. (1998)[1] with declinations δ > −25 degrees, right ascensions 12 < RA < 02 h and redshifts z < 0.175 that were classified as AGN on the basis of optical emission line diagnostic diagrams by Yuan et al. (2010)[2]. Of these 22 local ULIRGs with optical AGN activity, 19 (86%) were observed spectroscopically with the 4.2m William Herschel Telescope (WHT) as part of a study of the stellar populations in ULIRGs between 2005 and 2006[3]. 15 of the 19 were then identified as showing evidence for AGN-induced outflows on the basis of broad and/or blueshifted emission lines[4], and were subsequently re-observed with deeper spectroscopic observations using the ESO VLT telescope or the WHT telescope between 2013 and 2015, in order to better characterise the properties of their outflows. The full sample of 15 ULIRGs with repeat spectroscopic observations is representative (68% complete) of local ULIRGs that show evidence for optical AGN activity and associated warm gas outflows.

## 2. Evidence for an AGN in F01004-2237

The main evidence for an AGN in F01004-2237 is provided by optical and mid-IR spectroscopy observations. Based on optical diagnostic line ratio diagrams, this object has been classified as both an HII galaxy[5] and a Seyfert 2 galaxy[2]. The probable reason for this apparent ambiguity is that the optical spectrum of the nuclear regions is composite: while the intermediate and broad (500 < FWHM < 1700 km s$^{-1}$), blue-shifted (-1050 < Δv < -400 km s$^{-1}$) components to the emission lines have line ratios consistent with photoionization by an AGN, the ratios of the narrow components (FWHM < 200 km s$^{-1}$) are more consistent with photoionization by hot stars, or a combination of hot stars and an AGN[4]. Based on a variety of mid-IR continuum and emission line diagnostics, the AGN contribution to the bolometric luminosity of the source is estimated to be in the range 29 – 85%[6].

Despite the evidence for AGN activity from the optical and mid-IR emission lines, there is little clear evidence for luminous AGN activity in this source in other parts of the electromagnetic spectrum. At radio wavelengths, the 1.4 GHz radio power of the source is consistent with emission from a starburst and does not require the presence of a powerful relativistic jet component[4], while at X-ray wavelengths, the most recent X-ray study with the longest Chandra integration times found that the X-ray spectrum is consistent with a starburst origin, with no clear evidence for an AGN power-law[7] component. The latter result has been interpreted[7] in terms of the AGN being Compton-thick, with no radiation escaping the nuclear regions below an energy of 10 keV, due to heavy obscuration by circum-nuclear material. However, an alternative explanation is that the AGN activity is highly intermittent, and that at the observation epoch (December 2009 for the latest Chandra observations) the AGN was in a low activity state, such that the strong X-ray continuum emission and optical broad emission lines characteristic of a type 1 AGN fell below the detection limit. For this explanation to work, the low activity state could not have lasted longer than the light crossing time of the narrow-line region, else the high ionization narrow lines would also have switched off.

**3. Comparison with changing-look AGN**

There have been several reports in the literature of so-called changing-look AGN, in which the X-ray and/or optical emission of the AGN has decreased or increased by a substantial factor over a period of a decade or less, sometimes leading to the re-classification of the optical spectrum of the source (e.g. from a type 1 AGN to a type 2 AGN or vica versa). Concentrating on the changing-look objects in which the fluxes of AGN continuum and broad Balmer lines increased by a large factor and an optical spectrum was taken in the high state – the case most relevant to F01004-2237 – we have found reports of 13 such objects in the literature: NGC3516[8], Mrk1018[9], Mrk993[10], NGC1097[11], NGC7582[12], NGC3065[13], NGC2617[14], Mrk509[15], SDSS J233317[16], SDSS J214613[16], SDSS J225240[16], SDSS J002311[16] and HE 1136-2304[17]. These objects span the full range of AGN activity, from AGN with low ionization nuclear emission line regions (LINERS: NGC1097[11], NGC3065[13]) to luminous quasars (SDSS J233317[16], SDSS J214613[16], SDSS J225240[16], SDSS J002311[16]). However, unlike F01004-2237, none of these objects shows HeII and HeI lines that are unusually strong relative to Hβ in their high state spectra. It is also notable that the cause of the changing-look AGN phenomenon is uncertain. Indeed, it has been argued that the variability observed in some changing-look AGN may be due to TDEs[18].

**4. Host galaxy and stellar populations**

The single nucleus, compact appearance of F01004-2237 in optical and near-IR ground-based images[19] suggests that we are observing it at, or just after, the peak of the merger that has triggered the star formation activity. Although in higher resolution HST optical images[20] the barely-resolved compact nucleus dominates the emission, star clusters are visible in an arc structure at a radial distance of ~1kpc to the NE of the nucleus.

Based on colour-magnitude diagrams derived from HST optical images, Surace et al. (1998)[20] measure young ages ($t_{ysp} < 10$ Myr) for both the star clusters and the compact nucleus, and a stellar mass of $\sim 3 \times 10^8$ $M_\odot$ for the compact nucleus. Young ages for the nuclear stellar population are also indicated by our own spectral synthesis modelling of the continuum of the blue HST/STIS spectrum using the STARLIGHT code ($t_{ysp} \sim 8$ Myr), by the spectral synthesis modeling of the integrated light of the galaxy over a 5kpc aperture of Rodríguez Zaurín et al. (2009: $6 < t_{ysp} < 30$ Myr)[3], and by the detection of Wolf-Rayet features in the nuclear spectrum[21].

From the similar fluxes of the HeII and NIII features in the blend at 4660Å detected in the pre-2010 spectra, and the lack of strong NIVλ4604, HeIIλ5411 and CIVλ5808 emission, we deduce that the population of Wolf-Rayet stars is dominated by late WN stars, most likely of type WN9. A late WN classification is consistent with the earlier work of Armus et al. (1988)[21]. However, in order to adequately model the profile of the 4660Å blend in our HST/STIS spectrum, we also require a small contribution from WC7-8 stars, which emit a broad CIII/CIV feature at ~4650Å. Our best fit model for the 4460Å blend comprises a combination of WN7-9 and WC7-8 templates, along with a blueshifted (-1050 km s$^{-1}$) HeII line emitted by the outflow in the AGN NLR, with these components contributing 75, 18 and 7% respectively to the total flux in the blend (see Supplementary Figure 1(a)). The integrated flux of the 4660Å blend measured from the 2000 HST/STIS spectrum is 1.15x10$^{-15}$ erg s$^{-1}$ cm$^{-2}$. For the luminosity distance of F01004-2237 (523 Mpc) the total luminosities of the 4660Å features emitted by the WN7-9 and WC7-8 stars are 3.2x10$^{40}$ erg s$^{-1}$ and 7.9x10$^{39}$ erg s$^{-1}$ respectively. Assuming that single WN7-9 and WC7-8 stars both have $L_{4660}$ = 1.4x10$^{36}$ erg s$^{-1}$,[22-25] we deduce that there are ~23,000 WN7-9 and ~6,000 WC7-8 stars in the nucleus of F01004-2237 or ~3x10$^4$ WR stars in total. However, the number of Wolf-Rayet stars would be larger if the stellar population were significantly reddened.

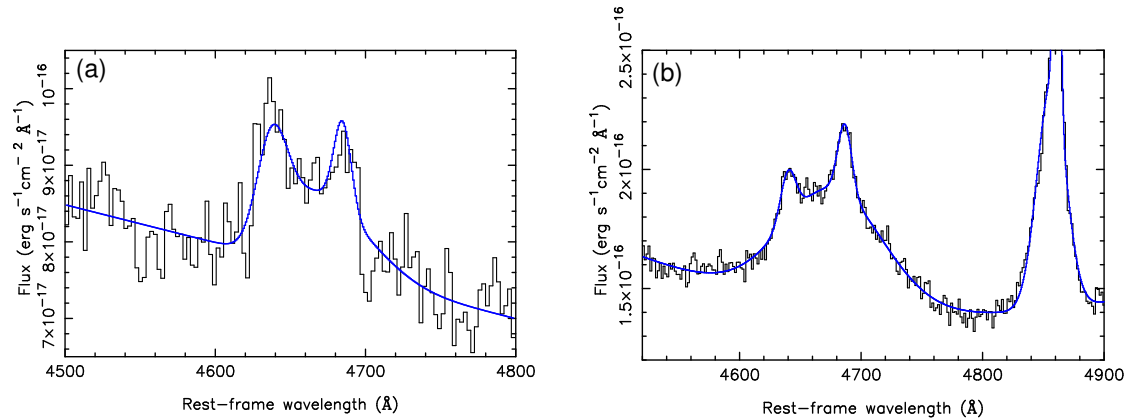

**Supplementary Figure 1.** Fits to the 4660Å feature in (a) the HST/STIS spectrum from 2000, and (b) the WHT/ISIS spectrum from 2015. Whereas the 4660Å feature in the 2000 HST/STIS spectrum can be fit with a combination of WN7-9 and WC7-8 Wolf-Rayet templates plus a blueshifted HeIIλ4686 line from the NLR (contributing 75, 18 and 7% respectively), the fit to the feature in the 2015 WHT/ISIS spectrum requires an additional broad HeIIλ4686 Gaussian component with FWHM~6200 km s$^{-1}$. Note that the broad HeIIλ4686 is not present in the Hβ line (visible to the right of (b)), which can be fit using a combination of two Gaussians with FWHM <200 km s$^{-1}$ and FWHM~1700 km s$^{-1}$.

## 5. Optical spectra and line identifications

Optical spectra were taken for F01004-2237 at 5 epochs prior to 2009: between 1985 and 1987 using the CTIO 4m telescope[21]; between 1994 and 1995 using the Kitt Peak 2.1m telescope[5]; in September 2000 using the Hubble Space Telescope (HST) with the STIS spectrograph[26]; in 2003 using the 6DF instrument; and in July 2005 using the ISIS spectrograph on the William Herschel Telescope[3]. None of these spectra show the strong, broad helium features present in the 2015 spectrum (e.g. see Supplementary Figure 2 for comparison of 2000, 2003, 2005 and 2015 spectra). However, since the spectra were taken in a variety of seeing conditions with different spectrograph apertures, it is important to be cautious about this comparison. In particular, the 1985/87, 1994/95, 2003 and 2005 spectra were taken in relatively poor (>2" FWHM) or uncertain seeing conditions with slit widths or optical fibre diameters ranging from 1.5 to 3.0". On the one hand, poor seeing can reduce the strength of spectral features emitted by the compact nucleus relative to those emitted by the more extended regions of the galaxy; while on the other, larger apertures admit more of the extended emission surrounding the nucleus.

We base our main comparison of the pre- and post-flare spectra of F01004-2237 on the 2000 HST/STIS and the 2015 WHT/ISIS spectra (see Figure 1 of the main text) because both of these sets of spectra suffer similar losses of flux from the compact nucleus due to resolution or seeing effects and finite slit widths. The details of these two sets of spectral observations and the aperture sizes used to extract the 1D spectra from the 2D long-slit spectra are shown in Supplementary Table 1. To estimate the slit losses for the WHT/ISIS data, we used the spectrum of a telluric standard star taken at the midpoint of the F01004-2237 observations with the same slit width, at similar airmass, and extracted from the long-slit data using the same extraction aperture (2.0x1.5"). Comparing the fluxes in the latter spectrum with published magnitudes for the star, we estimate that the slit losses for a point source in WHT/ISIS spectroscopic slit were 26±3% at the central wavelength of the R band (~6440Å). This is similar to the ~10-20% slit losses at a similar wavelength expected for the 0.2x0.2" extraction aperture used for the HST/STIS spectrum[27]. Therefore, for broader emission line components emitted by the compact nucleus, we can be confident about the accuracy of the comparison between the 2000 HST/STIS and 2015 WHT/ISIS spectra. However, the aperture used for the WHT/ISIS spectrum has 75x the area of that used for the HST/STIS spectrum, so it inevitably includes more emission from the extended regions surrounding the nucleus, which tend to emit relatively narrow emission lines (FWHM < 200 km s$^{-1}$)

| Telescope/Instrument | Date | Gratings | Spectral Res. (Å) | Exposure Time (s) | Extraction Aperture |
|---|---|---|---|---|---|
| HST/STIS | 02/09/2000 | G430L | 4.1 | 2895 | 0.2x0.2" |
| " | " | G750L | 7.4 | 1754 | " |
| WHT/ISIS | 15/09/2015 | R300B | 5.7 | 5400 | 1.5x2.0" |
| " | " | R316R | 5.5 | 5400 | " |

**Supplementary Table 1**. Observational details and extraction apertures used for the 2000 HST/STIS and 2015 WHT/ISIS spectra.

The HST/STIS spectra contain many of the emission lines commonly detected in AGN, including [NeV]λ3426, [OII]λ3727, [NeIII]λ3869, Hδ, Hβ, [OIII]λλ5007, 4959,4363, [OI]λ6300, Hα+[NII]λλ6548,6584, [SII]λλ6717,6731, along with the broad feature around ~4660Å that is attributed to Wolf-Rayet stars[21]. Although Farah et al. (2005)[26] claim that the blueshifted components to the [OIIII] lines detected in ground-based optical spectra are not detected in the STIS/HST spectrum, Gaussian fits to the [OIII] lines in the STIS/HST spectrum show that, in fact, they have similar central wavelengths and widths to the blueshifted components detected in the ground-based spectra. Rather, it is the *narrow* components to the [OIII] lines that are not detected in the STIS/HST spectrum, as might be expected given the small nuclear aperture used for this spectrum.

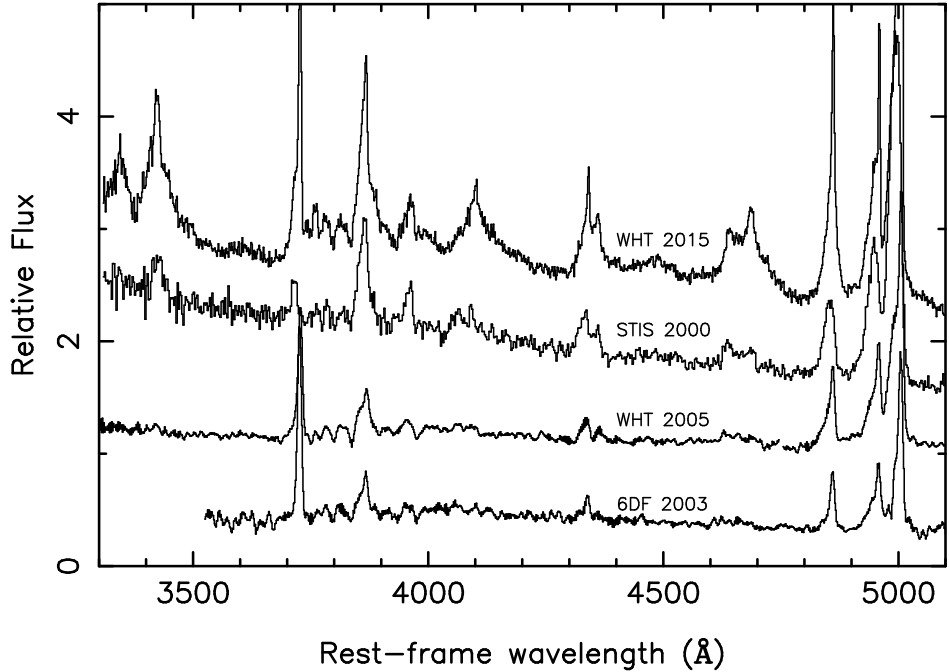

**Supplementary Figure 2.** Comparison of the spectra of the nuclear regions of F01004-2237 taken in 2000, 2003, 2005 and 2015. The relative fluxes are in wavelength units. In order to aid comparison the spectra have been shifted vertically by different amounts. Note that the accuracy of the relative flux calibration in the case of the 6DF spectrum is uncertain, but it is nonetheless possible to use this spectrum to gain an idea of the relative strengths of the various emission lines.

The 2015 WHT/ISIS spectrum shows some similarities with the HST/STIS spectrum in terms of the features detected, but it is markedly different in detail. Differences include the following.

- Stronger narrow components (FWHM < 200 km s$^{-1}$) to some emission features (e.g. [OII], H$\beta$, [OIII]$\lambda\lambda$5007,4959) in the 2015 spectrum, as expected given the much larger aperture size of the WHT/ISIS observations.
- The feature at 4660Å – attributed to a blend on NIII and HeII emission lines – is stronger by a factor of 5.6±1.1, and has broad wings that extend ~5,000 km s$^{-1}$ to the red of the centroid of the narrow HeII$\lambda$4686 line in the 2015 spectrum that are not visible in the adjacent H$\beta$ line (see Supplementary Figure 1(b) above). Although we identify the broadest emission line component of the blend with HeII$\lambda$4686, we cannot rule out a contribution from broad NIII$\lambda$4640.
- The HeI$\lambda$5876 feature is stronger (by a factor 3.7±0.2) in the 2015 spectrum where it is also notably broader than the [OI]$\lambda$6300 line. Unusually for a type 2 AGN, further HeI emission features are detected at 3889, 4471, 6678 and 7065Å in the 2015 spectrum.
- A broad feature is detected at ~4100Å in the 2015 spectrum, which is likely to be be a blend of H$\delta$, [SII] and HeII emission features, with a possible additional contribution from NIII.
- A broad feature is detected in the red wing of the [OI]$\lambda$6363 line in the 2015 spectrum, which we tentatively identify as [FeX]$\lambda$6374.

- The feature around 3426Å appears notably broader and stronger in 2015 than in 2000. If this feature is attributed entirely to [NeV], then its strength in 2015 is surprising given that strong [NeV] emission in AGN spectra is generally accompanied by strong [FeVII]λλ6087,5721 lines, which have a similar critical density and ionization potential. However, the [FeVII] lines are barely detected in the 2015 spectrum. It is possible that other, as yet unidentified, species contribute to the 3426Å feature.

| Line ID | Lab λ(Å) | FWHM (km s$^{-1}$) | Flux/10$^{-15}$ (erg cm$^{-2}$ s$^{-1}$) | (WHT2015)/ (HST2000) |
|---|---|---|---|---|
| [NeV] i | 3426 | 847 | 0.71±0.13 | |
| b | | 5296 | 4.54±0.24 | |
| i+b | | | 5,25±0.27 | 6.1±0.2 |
| [NeIII] n | 3869 | <200 | 0.44±0.07 | |
| b1 | | 1587 | 2.66±0.17 | |
| b2 | | 1430 | 0.04±0.09 | |
| b1+b2 | | | 2.70±0.19 | 1.2±0.2 |
| NIII i* | 4640 | 905 | 0.36±0.05 | |
| HeII i* | 4686 | 798 | 0.46±0.05 | |
| b | | 6181 | 5.58±0.27 | |
| NIII+HeII (tot) | | | 6.40±0.28 | 5.6±1.1 |
| Hβ n | 5861 | <200 | 1.11±0.04 | |
| b | | 1660 | 3.09±0.07 | 1.52±0.12 |
| [OIII]n | 5007 | <200 | 2.82±0.08 | |
| i | | 678 | 2.66±0.29 | |
| b | | 1676 | 6.12±0.31 | |
| i+b | | | 8.78±0.42 | 0.84±0.07 |
| HeI i* | 5876 | 598 | 0.46±0.06 | |
| b | | 3034 | 1.42±0.11 | |
| i+b | | | 1.88±0.13 | 3.7±0.2 |
| [OI] n | 6300 | 150 | 0.25±0.03 | |
| i1 | | 542 | 0.09±0.02 | |
| i2 | | 915 | 0.37±0.06 | |
| i1+i2 | | | 0.46±0.06 | 0.74±0.15 |

**Supplementary Table 2.** The results of Gaussian fits to the profiles of various emission features detected in the 2010 WHT/ISIS spectrum of F01004-2237 (columns 3 and 4). All the features require more than one Gaussian component for an acceptable fit. The emission lines are identified in columns 1 and 2, with narrow (FWHM < 500 km s$^{-1}$), intermediate (500 < FWHM < 1000 km s$^{-1}$) and broad (FWHM > 1000 km s$^{-1}$) components identified using "n", "I" and "b" labels. Features that may have a contribution from Wolf-Rayet stars are identified with asterisks. The final column shows the ratios between the 2010 WHT/ISIS and the 2000 HST/STIS emission line fluxes. Note that the latter ratios were estimated using the sums of the fluxes of the intermediate and broad components only, since these components are ,detected, and share similar kinematics, in both sets of spectra; the narrow emission line components are weak or absent in the narrower-slit HST/STIS spectra. The uncertainties in the emission line fluxes and the flux ratios are based entirely on the errors in the Gaussian fits and do not take into account uncertainties in the slit losses or systematic errors in the flux calibration.

Given that the narrower emission line features detected in the WHT/ISIS spectrum are likely to be emitted by star forming regions surrounding the nucleus[4], we concentrate our analysis on the broader emission line components. The fluxes and widths measured using multiple Gaussian fits to the emission lines are listed in Supplementary Table 2. The ratios of the fluxes of the broader emission line components measured in the 2015 WHT/ISIS spectra to those measured in the HST/STIS spectra are shown in column 5 of this table. Reassuringly, the fluxes of the broader components of the [NeIII], [OIII] and [OI] forbidden lines are consistent within 25% between the two epochs; given the uncertainties, there is no evidence for significant variability between the two epochs for these lines. This is consistent with the idea that the warm outflow in the nucleus of F01004-2237 is relatively compact and that the slit losses for the two epochs are similar.

**Supplementary information references**